\documentclass{article}
\usepackage{amsfonts,enumerate,amsmath,amssymb,marvosym}
\usepackage {graphicx}

\def\R{{\mathbb R}}
\def\Z{{\mathbb Z}}

\begin{document}
\title{Topological characteristics of oil and gas reservoirs and their applications}
\author{V.A. Baikov \thanks{Ufa State Aviation Technical University, 450025 Ufa, Russia, e-mail:baikov@ufanipi.ru}
\and
R.R. Gilmanov \thanks{OOO ``Gazpromneft NTC'', 190000, St. Petersburg, Russia, e-mail: Gilmanov.RR@gazpromneft-ntc.ru}
\and
I.A. Taimanov \thanks{Sobolev Institute of Mathematics, 630090, Novosibirsk, Russia, and Novosibirsk State University, 630090 Novosibirsk, Russia,  e-mail: taimanov@math.nsc.ru}
\and
A.A. Yakovlev \thanks{OOO ``Gazpromneft NTC'', 190000, St. Petersburg, Russia, e-mail: Yakovlev.AAle@gazpromneft-ntc.ru}
}
\maketitle


At present no company develops oil and gas fields without constructing geologic and hydrodynamic models. This is due in particular to the fact that recently the  emphasis in design, planning and monitoring has shifted to over-dissected and low-permeability reservoirs. To assess economic efficiency and optimal placement of wells and to predict hydrocarbon production levels, it is important to have some quantitative representation  of the object under study. This requires a mathematical measure of a geological description and mathematical models of the structure of oil and gas reservoirs.

The geological modeling based on digital oil and gas reservoirs splits into two parts:

\begin{enumerate}
\item
a digital interpolation of a reservoir based on the observed data and on the probabilistic nature of functions which describe formations;

\item
a hydrodynamic modeling based on the filtration equations.
\end{enumerate}

Therewith it is important to choose the most suitable model for developing.
In \cite{BBTY} we proposed to use topological characteristics of digital reservoirs as
one of the factors for choosing a model. These characteristics can be used for

\begin{itemize}
\item
{\sl comparing different stochastic realizations of the same reservoir and, in particular, using that information for choosing a certain realization for industrial development;}

\item
{\sl estimating the topological complexity of a reservoir.}
\end{itemize}

In particular, this method can help to choose realization that gives a reliable model of a reservoir and
whose exploration does not need resource-intensive  calculations.

For optimizing a process of geological and hydrodynamic modeling by limiting a series of direct problems there
arise tasks of creating a list of topological, geometric, fractal, and other characteristics of inhomogeneous anisotropic environment and their subsequent influences on the construction of a model. Such problems are studied in
geometry of random fields.

In \cite{BBTY} we demonstrated that two different stochastic approaches for constructing digital reservoirs gives
topologically similar pictures and one of them, being more rough is nevertheless preferable for
dynamical modeling due to its relative simplicity for numerical dynamical modeling.
We expose some results on the Betti numbers of reservoirs in \S 2.

 Since the  ``permeability'' function $Z$ determines a natural filtration of the reservoir by the excursion sets it is reasonable to pick up the topological picture of the filtration and use for that the persistent homology \cite{P1,P2}
 (see also \cite{EH2008,C2009,EH,F}). In this framework

 \begin{itemize}
 \item
{\sl  the ``bottleneck'' distance between persistent diagrams can be used for estimating differences between
 reservoirs and not only between their models.}
 \end{itemize}

We discuss this approach in \S 3.

\section{Stochastic and topological preliminaries}

\subsection{The kriging}

The digital reservoirs under consideration are constructed by the kriging method from the observed data
(see \cite{Matheron,D,B}  and the references therein).  This method has many variations,
based on the same idea, and we explain which one we use.

In our case a digital reservoir is a union of cubes such that a certain characteristic related to the permeability is
a function $Z$ on the set of these cubes. For simplicity we assume that the reservoir is the domain
$$
D = \{x_0 \leq x \leq x_0+ N_x\delta_x, y_0 \leq y \leq y_0 + N_y\delta_y, z_0 \leq z \leq z_0 +N_z\delta_z\},
$$
where $x$ and $y$ are the lateral coordinates and $z$ is the height coordinate, while
$\delta_x,\delta_y,\delta_z$ are the length, width and depth of the elementary cube.
The domain $D$ splits into $N_x N_y N_z$ elementary cubes $C_{k_x,k_y,k_z}$ defined by the inequalities
$$
x_0 + (k_x-1)\delta_x \leq x \leq x_0+ k_x\delta_x,  \ \ \
y_0 + (k_y-1)\delta_y \leq y \leq y_0 + k_y\delta_y,
$$
$$
z_0 + (k_z-1) \delta_z \leq z \leq z_0 +k_z\delta_z,
$$
where the triples $(k_x,k_y,k_z)$ parameterize the elementary cubes and
$Z$ is considered as a function on these triples:
$$
Z = Z(k_x,k_y,k_z).
$$
We denote the set of all these triples by
$$
S = \{1,\dots,N_x\} \times \{1,\dots,N_y\} \times \{1,\dots,N_z\}
$$
and for every subset $S^\prime \subset S$ we denote by $D_{S^\prime}$ the union of elementary cubes corresponding to
triples from $S^\prime$:
$$
D_{S^\prime} \subset D, \ \ \ D_S = D.
$$

Let the function $Z$ is known for some set $S^\prime$ of elementary cubes.
For instance, this may be the observed data from wells.
We extend $Z$ onto $S$ by the following stochastic regression method.

Let us choose a procedure for choosing randomly an element $p_{M+1}$ from
$S \setminus S^\prime$ where
$M$ is the number of elements of $S^\prime$.

The function $Z$ is considered as a random field such that

\begin{enumerate}
\item
it is stationary, i.e., it has the same expectations at all points:
$$
{\mathrm E}\,(Z(p)) = {\mathrm E}\,(Z(q)) = m \ \ \ \mbox{for all $p,q \in S$}
$$
and $m$ is known ({\it simple kriging});

\item
the correlation between two random variables depends only on the spatial distance between them:
$$
C(Z(p),Z(q)) = C(|p-q|).
$$
\end{enumerate}

Here we mean by  the spatial distance $|p-q|$ between elementary cubes the distance between their centers.
The correlators are given by the variogram:
$$
\gamma(|h|) = \frac{1}{2} {\mathrm E}\,((Z(p)-Z(p+h))^2) =
C(0)-C(|h|).
$$
This variogram is derived from observations.

Given a sample $(Z(p_1),\dots,Z(p_M))$, the values of $Z$ at $S^\prime$,
the value $s_{M+1}$ is obtained from the conditions
$$
Z^\ast = \sum_{i=1}^M \lambda_i Z(p_i), \ \ \ \sum \lambda_i = 1, \ \
{\mathrm E}\,((s_{M+1} - Z^\ast)^2) \to \min
$$
which results in the system:
$$
\sum \lambda_i = 1,
$$
$$
\sigma^2 = C(0) - 2\sum_i \lambda_i C(|p_i-p_0|) + \sum_{i,j} \lambda_i \lambda_j C(|p_i-p_j|) \to \min,
$$
where $\sigma^2$ is a measure of precision.
To determine $Z(p_{M+1})$ we put
$$
Z(p_{M+1}) = s_{M+1} + \xi
$$
where the random process $\xi$ satisfies the Gauss distribution with
${\mathrm E} = 0, \mathrm{Var}\, = \sigma^2$.
Thus we derive the new sample $Z(p_1),\dots,Z(p_{M+1})$, add $p_{M+1}$ to $S^\prime$ and
resume by the same way
until we extend $Z$ onto $S$.

This procedure is called {\it the sequential Gauss simulation} (SGS).
The standard variograms that are used are

\begin{itemize}
\item
the Gaussian variogram:
$C(h) = \mathrm{const}\,(1- e^{-h^2/R^2})$,

\item
the exponential variogram: $C(h) = \mathrm{const}\,(1 - e^{-h/R})$.
\end{itemize}

In both cases $R$ is the radius of a variogram.

There is another stochastic regression in which the field is represented as a linear combination of the first $M$
Legendre polynomials
$$
Z(x,y,h) = \sum_{i=1}^M a_i(x,y) L_i(h)
$$
where $x$ and $y$ are the lateral variables, $h$ is the depth,
and $a_i(x,y)$ are independent random fields which are extrapolated by some two-dimensional stochastic regression.
This method is called the {\it spectral expansion}.

In developing oil formations an important data is
$$
Z(p) = \mathrm{GL}(p)
$$
which is the gamma logging, i.e., the natural radioactivity of formation.
We put
$$
\alpha (p) = \frac{\mathrm{GL}(p) - \mathrm{GL}_{\min}}{\mathrm{GL}_{\max} - \mathrm{GL}_{\min}}
$$
and assume that $p$ belongs to the formation if
$$
\alpha (p) \leq \alpha_0,
$$
where $\alpha_0$ is the excursion coefficient.
By varying $\alpha_0$ we obtain a filtration of the reservoir $D$ by the excursion sets
$$
D_1 \subset D_2 \subset \dots \subset D_M \subset D_\infty = D
$$
where $\varepsilon_1 < \varepsilon_2 < \dots < \varepsilon_M <1$
and $D_i = D_{S_{\varepsilon_i}}, S_{\varepsilon_i} = \{\alpha \leq \varepsilon_i\} \subset S$.

The double difference parameter $\alpha$
is widely used in practice. Therewith
$\mathrm{GL}_{\min}$ and $\mathrm{GL}_{\max}$ are the minimal and maximal values of $\mathrm{GL}$ which correspond to a neat oil and gas reservoir and a clay which supports a reservoir.
These values should not be confused with the absolute minimum and
maxima values of $\mathrm{GL}$ with which they coincide only in exceptional model
examples. This is due, in particular, to the possible presence of minor anomaly noises
and to effects of stochastic modeling. Moreover, often $\mathrm{GL}_{\min}$ and $\mathrm{GL}_{\max}$ are calibrated by certain samples. Therefore sometimes in modeling there appear values of
$\alpha$ which are less than zero or greater or equal than $1$ and, if not to take care of that,
$D_\infty$ may not coincide with  $D$.

The stochastic modeling leads to very adequate pictures of reservoirs. As an example, we present on Fig. 1
the model of a reservoir obtained from the observed data. This model is obtained by the SGS data, the
parameters of the domain $D$ are
$N_x = N_y = 120, N_z = 490, \delta_x=\delta_y = 50\, \mathrm{m}, \delta_z = 0.4\, \mathrm{m}$,
the colors vary from light to dark that corresponds to the variation of $\alpha$ from small to large values,
the reservoir corresponds to the excursion $\alpha=0.6$.

\begin{figure}[ht]
\centerline{\includegraphics[width=5in,height=2.8in]{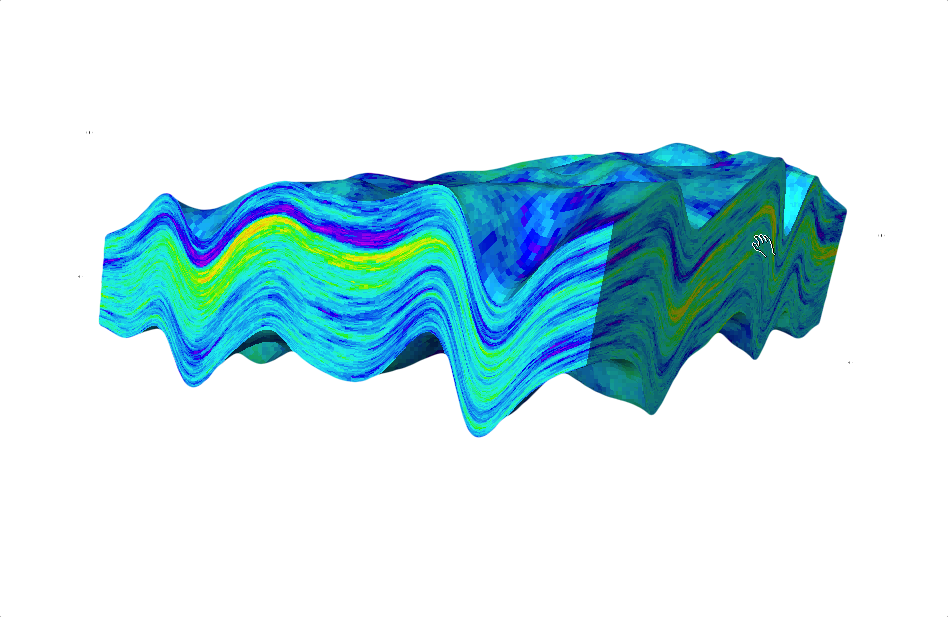}}
\caption{A reservoir modeled by the SGS method}
\label{fig1}
\end{figure}

\subsection{Topological characteristics of $3$-dimensional bodies}

We consider three-dimensional solid bodies as composed from elementary cubes (cubic complexes).

The main topological characteristics of such bodies are their Betti numbers (with $\Z_2$ coefficients)
$b_0$, $b_1$ and $b_2$. The meaning of these characteristics is very natural:
$b_0$ is the number of connected components, $b_1$ is the number of handles, and $b_2$ is
the number of holes (cavities).

If we start from a solid cube, remove $k$ holes from its interior and attach $l$ handles to the cube we obtain the body $X$ for which
$b_0 = 1,  b_1 = l, b_2 = k$.
The Betti numbers of a topological space are the ranks of the corresponding homology groups
$H_i(X;\Z_2)$
(here an in the sequel we consider the homology groups with coefficients with $\Z_2$
and for denote them by $H_i(X)$ for simplicity):
$$
H_0 = \Z_2^{b_0} = \Z_2 \oplus \dots \oplus \Z_2 \ \ \mbox{(the sum of $b_0$ copies of $\Z_2$)},
$$
$$
H_1 = \Z_2^{b_1}, \ \ \ H_2= \Z_2^{b_2}.
$$
The alternated sum
$$
\chi = b_0 - b_1 + b_2
$$
is called the Euler characteristic of a three-dimensional solid body.

We refer to \cite{EH} for an introductory exposition, of these topological characteristics, oriented to applications.
We recall that if two topological spaces (bodies) are topologically equivalent (or homeomorphic),
i.e. if there exists a continuous in both sides one-to-one correspondence between points of spaces, then
they have the same topological characteristics (the Betti numbers, homology groups etc.)

The Euler characteristic may be easily computed from the cubic decomposition of the solid body $X$.
We have $X$ as a union of cubes such that

a) two different cubes may intersect each other only by a joint vertex, edge or
face;

b) two different faces may intersect each other only by a  joint vertex or edge;

c) two different edges my intersect each other only by a joint vertex.

We denote by $c_0$ the number of vertices; by $c_1$ the number of edges; by $c_2$ the number of
faces; and by $c_3$ the number of cubes. For instance, the cubic decomposition of an elementary cube has
$8$ vertices, $12$ edges, $6$ faces and $1$ cube.

The Euler characteristic of a three-dimensional body is given also by the formula:
\begin{equation}
\label{euler}
\chi = c_0 - c_1 + c_2 - c_3.
\end{equation}
For an elementary cube we have $\chi = 8-12+6-1=1$.

If $X$ is body composed from finitely many cubes and
lies in the three-space $\R^3$, then the particular case of the Alexander duality implies that
$$
b_2(X) = b_0(\R^3 \setminus X) - 1.
$$
Hence, in difference with higher-dimensional space, for calculating the Betti numbers of
a three-dimensional body $X$ it is enough to find its Euler characteristic
$\chi$ from the cubic decomposition (see (\ref{euler})) and the numbers of connected components of $X$ and of its complement.
Then $b_1$ is given by the equality
\begin{equation}
\label{b1}
b_1(X) = b_0(X) + b_0(\R^3 \setminus X) - 1 -\chi(X).
\end{equation}
That drastically simplifies the calculation of the Betti numbers and reduces it to calculating of the numbers of
connected components of cubic complexes.

The development of numerical methods for calculating the Betti numbers is necessary because reservoirs may be
very complicated. In particular, some numerical  approach, based on a certain discretization of the Morse theory to finding the Betti numbers of reservoirs was exposed in \cite{BT}.

\section{The Betti numbers of digital reservoirs}

Since geological formations are natural examples of three-dimensional solid bodies, it is reasonable to consider their
topology for geological applications however that was started not long ago  (see \cite{BBTY} for oil and gas reservoirs and \cite{A1,A2} and references therein for applications to structural geology).

Let us demonstrate numerical examples of the Betti numbers of reservoirs.

Given the excursion parameter $\alpha_0$, we have the cubic complex
$$
D_{\alpha_0} = \cup_{Z(k_x,k_y,k_z) \leq \alpha_0} C_{k_x,k_y,k_z}
$$
composed from all cubes for which $Z \leq \alpha_0$.

To construct from $D_{\alpha_0}$ the topological model of the corresponding reservoir we have to keep in mind that
if two cubes do have only a joint edge or a joint vertex then there is no percolation between them through
the joint cell (edge or vertex). The percolation between two adjacent cubes is possible only
through a joint two-dimensional face.
Hence we have to unstack all such cubes and obtain an abstract cubic complex $X_{\alpha_0}$.
This complex is the right model that respects the percolation rules and can be chosen for industrial development.

To give an impression on the topological complexity of reservoirs we present results of some calculations
 corresponding to simulated reservoirs (see Table 1). We consider the four digital models that correspond to
 the exponential variogram $C(h) = (1-e^{-h/R})$ with $R=500\, \mathrm{m}$ and $R=1000\, \mathrm{m}$ and
 to the Gaussian variogram $C(h) = (1-e^{-h^2/R^2})$ with $R=500\, \mathrm{m}$ and $R=1000\, \mathrm{m}$. The data of the reservoirs are $N_x=N_y=N_z=100, \delta_x = \delta_y = 100\, \mathrm{m}, \delta_z = 1\, \mathrm{m}$.

\begin{table}[htbp]
\caption{The Betti numbers and the Euler characteristic}
\begin{tabular}
{|p{40pt}|p{50pt}|p{50pt}|p{50pt}|p{50pt}|}
\hline
$\alpha_0$&
b$_{0}$ &
b$_{1}$ &
b$_{2}$ &
$\chi $\\
\hline
\raisebox{-1.50ex}[0cm][0cm]{0.1}&
1664 &
0 &
0 &
1664\\
\cline{2-5}
 &
477&
1 &
0&
476\\
\cline{2-5}
& 5042&
1 &
0 &
5041\\
\cline{2-5}
& 3491&
2  &
0 &
3489\\
\hline
\raisebox{-1.50ex}[0cm][0cm]{0.2}&
4751&
9 &
0 &
4742 \\
\cline{2-5}
 &
1330 &
10 &
0&
1320\\
\cline{2-5}
& 18691&
9 &
0 &
18682 \\
\cline{2-5}
& 11779 &
60 &
0 &
11719 \\
\hline
\raisebox{-1.50ex}[0cm][0cm]{0.3}&
6113&
260 &
0 &
5853\\
\cline{2-5}
 &
1606 &
110 &
0&
1496\\
\cline{2-5}
& 32601&
495 &
3 &
32109 \\
\cline{2-5}
& 18757&
997 &
12 &
17772\\
\hline
\raisebox{-1.50ex}[0cm][0cm]{0.4}&
1932&
3682 &
3 &
-1747\\
\cline{2-5}
 &
487&
1150&
0&
-663\\
\cline{2-5}
& 20905&
9355 &
329 &
11879\\
\cline{2-5}
& 12813 &
9455&
391 &
3749\\
\hline
\raisebox{-1.50ex}[0cm][0cm]{0.5}&
245&
11389&
163 &
-10981\\
\cline{2-5}
 &
55&
2995&
29&
-2911\\
\cline{2-5}
& 4971&
45256 &
4187&
-36098\\
\cline{2-5}
& 3713&
28695&
3324 &
-21658\\
\hline
\raisebox{-1.50ex}[0cm][0cm]{0.6}&
18&
8523&
1434 &
-7071\\
\cline{2-5}
 &
1 &
1927&
265 &
-1661\\
\cline{2-5}
& 528&
53806&
18129 &
-35149\\
\cline{2-5}
& 473&
29870&
11421 &
-17976\\
\hline
\raisebox{-1.50ex}[0cm][0cm]{0.7}&
1&
3133&
4903 &
1771\\
\cline{2-5}
 &
1&
545&
1045&
501\\
\cline{2-5}
& 14&
28988 &
29705&
731\\
\cline{2-5}
& 31&
15949&
16832&
914\\
\hline
\raisebox{-1.50ex}[0cm][0cm]{0.8}&
1&
721 &
4132 &
3412\\
\cline{2-5}
 &
1 &
92 &
974 &
883 \\
\cline{2-5}
& 1&
6658 &
17563 &
10906 \\
\cline{2-5}
& 3 &
4216 &
10220 &
6007 \\
\hline
\raisebox{-1.50ex}[0cm][0cm]{0.9}&
1&
85 &
1488  &
1404\\
\cline{2-5}
 &
1 &
6 &
389 &
384 \\
\cline{2-5}
& 1 &
637 &
4798 &
4162 \\
\cline{2-5}
& 1 &
608 &
3171 &
2564\\
\hline
\end{tabular}
\label{tab1}
\end{table}

The numerical experiment shows the stability of the integral topological characteristics (the Betti numbers weighted by a volume) under
stochastic modeling, sensitivity to the type and the rank of the variogram (see Figures 2 and 3). 
The characteristics lie on similar cycles and in both cases the inner (smaller) cycle corresponds to the largest value of $R (=1000)$.
Thus these characteristics can serve as classifiers
for assigning digital geological models to equivalent and as a consequence, they have important applied values for the determination of analogs in
the modeling of poorly studied oil fields (the case of lack of information for a reliable distribution of reservoir properties).

\begin{figure}[ht]
\centerline{\includegraphics[width=3in,height=3in]{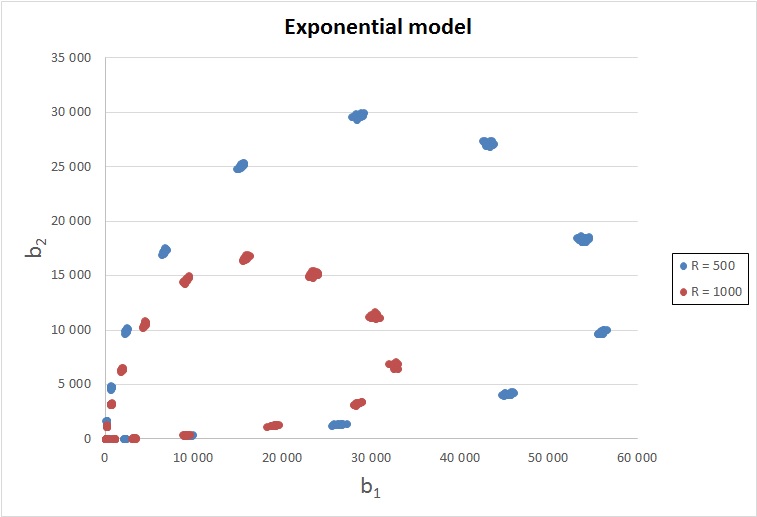}}
\caption{Relations between the weighted Betti numbers of digital reservoirs for the exponential variogram}
\end{figure}

\begin{figure}[ht]
\centerline{\includegraphics[width=3in,height=3in]{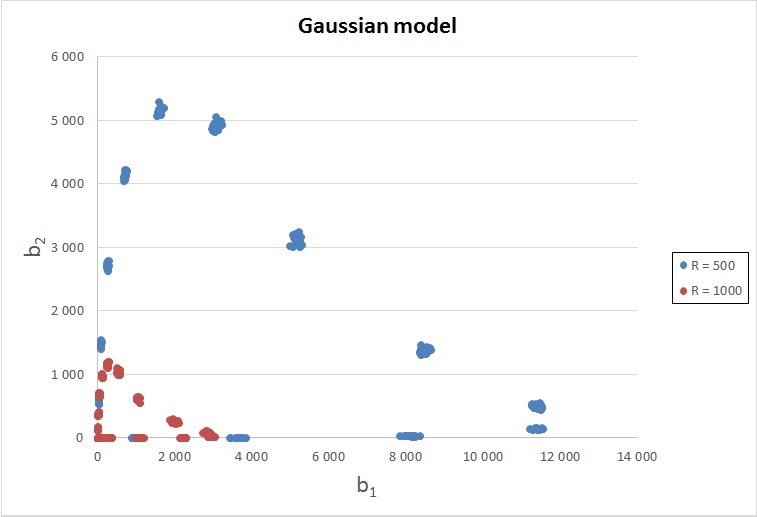}}
\caption{Relations between the weighted Betti numbers of digital reservoirs for the Gaussianl variogram}
\end{figure}

We remark that for regular (smooth) Gaussian fields the expectation of the Euler characteristic of the excursion set was found in \cite{Adler} and
this approach was extended for random fields related to Gaussian \cite{AT}. The formula for the expectation of the number of components, i.e. of $b_0$,
is not derived until recently, as well for the expectations of other relations between topological and metric characteristics
which are demonstrated in Figures 2 and 3.

\section{The ``bottleneck'' distance between digital reservoirs}

To every continuous mapping of topological spaces and, in particular, of three-dimensional bodies
$$
f: X \to Y
$$
there correspond the homomorphisms of their homology groups
$$
f_\ast: H_i(X) \to H_i(Y).
$$
Given a filtration
$$
X_1 \subset \dots \subset X_M
$$
where every inclusion is treated as the embedding $X_i \to X_{i+1}$ and the compositions of such inclusions give
embeddings $X_i \to X_j$ for all $i<j$, we have for every dimension $q$ the homomorphism
$$
f^{i,j}_q: H_q(X_i) \to H_q(X_j).
$$
The persistent homology groups \cite{P1,P2,EH2008}
are defined as
$$
H^{i,j}_q = \mathrm{Im}\, f^{i,j}_q = f_\ast(H_q(X_i)) \subset H_q(X_j).
$$

Let us fix $q$. To every generator $z \in H_q(X_i)$ such that $z$ does not lie in the image of $H_q(X_{i-1}) \to H_q(X_i)$, it is mapped into
nontrivial elements by homomorphisms $H_q(X_i) \to H_q(X_{j-1})$ and $f^{i,j}_q(z)=0$ we correspond a point on the plane with coordinates
$(i,j)$. Here we recall that we consider homology groups with coefficients in $\Z_2$ and this procedure is defined for all coefficients and
also for continuous values of indices $i$.

The persistent diagram of a filtration (for the $q$-dimensional homology) is the union $U$ of all such points taken with their multiplicities and points of
of the diagonal $(x,x) \subset \R^2$ taken with infinite multiplicities.

The persistent homology and their persistent diagrams play a fundamental role in the modern topological data analysis \cite{C2009,F}.

The persistent diagrams are stable under small perturbations of initial topological data \cite{CEH}. The distance between different persistent diagrams is given
by the bottleneck distance defined as follows
$$
\rho(U,V) = \inf_{\eta:U \to V} \sup_{u\in U} |u - \eta(u)|,
$$
where the infimum is taken over all bijections $\eta: U \to V$ and the norm $|u|$ on the plane has the form $|u| = |x|+|y|$ where $u=(x,y)$.
This bottleneck distance plays an important role in the optimization theory and different algorithms for its computation were recently
used in topological data analysis (see, for instance, \cite{EAK,KMN}).

The bottleneck distance can be used for comparing different digital reservoirs. We present results of some numerical experiments.
We compute the bottleneck distances between the $0$-dimensional ($q=0$) persistent diagrams corresponding to $8$ digital reservoirs which splits into four pairs corresponding to the exponential ($E$) and
Gaussian ($G$) variograms and to $R=500 \, \mathrm{m}$ or $R=1000 \, \mathrm{m}$. This reservoirs correspond to $N_x=N_y=N_z = 25$,
$\delta_x=\delta_y = 400\,\mathrm{m}, \delta_z = 4\, \mathrm{m}$, and the step of the discretized excursion parameter $\alpha_0$ (the step of the excursion filtration) is equal to $\Delta \alpha_0 = 0.01$.
Metrically these reservoirs have the same form --- $10000 \,\mathrm{m} \times 10000 \,\mathrm{m} \times 400 \,\mathrm{m}$ --- as the reservoirs
in Table 1. But we consider a rough decomposition because the complexity of the calculation of the bottleneck distance is $O(n^2 \log n)$ where
$n$ is the number of points in the persistence diagram and for some digital reservoirs from  Table 1 we have $n \approx 50000$ which makes the calculation time- and resource-consuming. Keeping in mind that $0 \leq \alpha \leq 1$ and hence the distance between such diagrams is at most $1$, the data shows that this metric
really distinguishes diagrams but it needs to understand for which types of digital reservoirs and, in particular, for which ratios of $R$ and the sizes of elementary cubes
this approach gives applicable answers.

\begin{table}[htbp]
\caption{The bottleneck distance}
\begin{tabular}
{|p{45pt}|p{30pt}|p{30pt}|p{30pt}|p{30pt}|p{30pt}|p{30pt}|p{30pt}|p{30pt}|}
\hline
& E500-1 & E500-2 & E1000-1 & E1000-2 & G500-1 & G500-2 & G1000-1 & G1000-2 \\
\hline
E500-1 & 0 &  0.11& 0.11 & 0.12 & 0.13 & 0.09 & 0.15 & 0.16 \\
E500-2 & 0.11 & 0 & 0.055 & 0.1 & 0.07 & 0.06 & 0.11 & 0.13 \\
E1000-1 & 0.11 & 0.055 & 0 & 0.09 & 0.05 & 0.06 & 0.11 & 0.11 \\
E1000-2 & 0.12 & 0.1 & 0.09 & 0 & 0.05 & 0.05 & 0.07 & 0.07 \\
G500-1 & 0.13 & 0.07 & 0.05 & 0.05 & 0 & 0.07 & 0.08  & 0.08\\
G500-2 & 0.09 & 0.06 & 0.06 & 0.05 & 0.07 & 0 & 0.08 & 0.09 \\
G1000-1& 0.15 & 0.11 & 0.11 & 0.07 & 0.08 & 0.08 & 0 & 0.05 \\
G1000-2& 0.16 & 0.13 & 0.11 & 0.07 & 0.08 & 0.09 & 0.05 & 0 \\
\hline
\end{tabular}
\label{tab2}
\end{table}


\begin{thebibliography}{MMM}
\bibitem{BBTY}
Bazaikin, Ya.V., Baikov, V.A., Taimanov, I.A., Yakovlev, A.A.:
Numerical analysis of topological characteristics of three-dimensional geological models of oil and gas fields.
Mathematical Modeling {\bf 25}:10, 19--31 (2013). (Russian)


\bibitem{P1}
Edelsbrunner, H., Letscher, D., Zomorodian, A.:
Topological persistence and simplification.
Discrete Comput. Geom. {\bf 28}, 511-533 (2002). doi: 10.1007/s00454-002-2885-2.

\bibitem{P2}
Zomorodian, A., Carlsson, G.:
Computing persistent homology.
Discrete Comput. Geom. {\bf 33}, 249--274 (2005). doi: 10.1007/s00454-004-1146-y.

\bibitem{EH2008}
Edelsbrunner, H., and Harer, J.:
Persistent homology --- a survey. In: Goodman, J.E., Pach, J., Pollack R. (eds.) 
Surveys on discrete and computational geometry, 
Contemp. Math., vol. 453, pp. 257--282. Amer. Math. Soc., Providence, RI (2008).

\bibitem{C2009}
Carlsson, G.:
Topology and data.
Bull. Amer. Math. Soc. (N.S.) {\bf 46}, 255--308 (2009). doi: 10.1090/S0273-0979-09-01249-X.

\bibitem{EH}
Edelsbrunner, H., Harer, J.L.:
Computational Topology.
An Introduction. American Mathematical Society, Providence, RI (2010).

\bibitem{F}
Ferri, M.:
Persistent topology for natural data analysis - A survey. In: https://arxiv.org/abs/1706.00411.

\bibitem{Matheron}
Matheron, G.:
Trait\'e de Geostatistique Appliqu\'ee. Editions BGRM, Paris (1962).

\bibitem{D}
Dubrule, O.:
Geostatistics in Petroleum Geology.
American Association of Petroleum Geologists, Tulsa (1998).

\bibitem{B}
Baikov, V.A., Bakirov, N.K., Yakovlev, A.A.:
Mathematical Geology.I. Introduction to Geostatistics.
Izhevsk Institute of Computer Sciences, Izhevsk (2012). (Russian)

\bibitem{BT}
Bazaikin, Ya.V., Taimanov, I.A.:
On a numerical algorithm for computing topological characteristics of three-dimensional bodies.
Journal of Computational Mathematics and Mathematical Physics {\bf 53}, 523--530 (2013).  (Russian)

\bibitem{A1}
Thiele, S.T., Jessel M.W., Lindsay, M., Ogarko, V., Wellmann, J.F., and Pakyuz--Charrier, E.:
The topology of geology 1: Topological analysis.
Journal of Structural Geology {\bf 91}, 27--38 (2016).
doi: 10.1016/j.jsg.2016.08.009.

\bibitem{A2}
Thiele, S.T., Jessel M.W., Lindsay, M., Wellmann, J.F., and Pakyuz--Charrier, E.:
The topology of geology 2: Topological uncertainty.
Journal of Structural Geology {\bf 91}, 74--87 (2016).
doi: 10.1016/j.jsg.2016.08.010.

\bibitem{Adler}
Adler, R.J.:
The Geometry of Random Fields, Wiley, London (1981).

\bibitem{AT}
Adler, R.J., Taylor, J.E:
Random Fields and Geometry.
Springer, Heidelberg (2007).

\bibitem{CEH}
Cohen-Steiner, D., Edelsbrunner, H., Harer, J.:
Stability of persistence diagrams. 
Discrete and Computational Geometry {\bf 37}, 103--120 (2007). doi: 10.1007/s00454-006-1276-5.

\bibitem{EAK}
Efrat,  A., Itai, A., Katz, M.J.:
Geometry  helps  in  bottleneck  matching  and related problems.
Algorithmica {\bf 31}:1 (2001), 1--28. doi: 10.1007/s00453-001-0016-8.

\bibitem{KMN}
Kerber, M., Morozov, D., Nigmetov, A.:
Geometry helps to compare persistence diagrams.
In: https://arxiv.org/abs/1606.03357.




\end{thebibliography}
\end{document}